\documentclass[prl,showpacs,floatfix,superscriptaddress,twocolumn]{revtex4-2}  
\usepackage{amsfonts}
\usepackage{amsmath}
\usepackage{graphics}
\usepackage{epsfig}
\usepackage{amssymb}
\usepackage{color}
\usepackage{bm}

\newcommand{\cH}{{\cal H}}

\newcommand{\cS}{\mathcal{S}}

\newcommand{\bPsi}{\mathbf{\Psi }}

\newcommand{\bpsi}{ \mbox{\boldmath$\psi$\unboldmath}  }

\newcommand{\bvarphi}{ \mbox{\boldmath$\varphi$\unboldmath}  }

\newcommand{\bu}{{\bm u}}

\newcommand{\bS}{{\bm S}}

\begin{document}

\title{Quantized transport of solitons in Bose-Einstein condensates driven by spin-orbit coupling}

\author{Yaroslav V. Kartashov}
\affiliation{Institute of Spectroscopy, Russian Academy of Sciences, Troitsk, Moscow, 108840, Russia}

\author{Vladimir V. Konotop}
\affiliation{ 
 Centro de F\'isica Te\'orica e Computacional Universidade de Lisboa, Campo Grande 2, Edif\'icio C8, Lisboa 1749-016, Portugal} 
\affiliation{Departamento de F\'isica, Faculdade de Ci\^encias,  Universidade de Lisboa, Campo Grande 2, Edif\'icio C8, Lisboa 1749-016, Portugal
}

\author{Dmitry A. Zezyulin}
\affiliation{School of Physics and Engineering, ITMO University, St. Petersburg 197101, Russia}

\date{\today}

\begin{abstract}

We demonstrate that linear and nonlinear Thouless pumping can be realized in two-component elongated Bose-Einstein condensates using helicoidal spin-orbit coupling that slides with respect to a static optical lattice, identical for both spinor components. Stable quantized transport is found for solitons in semi-infinite and finite gaps, within certain intervals of chemical potentials and numbers of atoms. In the semi-infinite gap, the transport is arrested for solitons with sufficiently large number of atoms. We elucidate the important role of Zeeman splitting in the control of quantized transport, which disappears when the longitudinal component of the Zeeman field is removed.

\end{abstract}

\maketitle
 
\paragraph{{Introduction.}}  Thouless pumping \cite{Thouless} is a fundamental phenomenon that involves quantized, typically adiabatic, transport of a physical quantity through a dynamically modulated periodic medium. It  has attracted considerable attention and has been observed in diverse physical systems, including atomic systems~\cite{NaToTa2016, TaCoFa2017, LoSHa2018, LosZilAlBlo2020, NaTaKe2021}, arrays of optical waveguides~\cite{ZilHuaJo2018,Cerjan2020} and optically induced lattices \cite{Wang2022, Yanga2024}  {(see the review~\cite{Citro2023} for further references)}. Due to inherent nonlinearity of the above platforms, there has been growing interest in topological pumping of solitons. This interest was sparked by recent experimental observation of Thouless pumping of solitons in optical waveguides \cite{JuMuRe2021, Jurgensen2023} and theoretical studies of this phenomenon in optical \cite{Jurgensen2022, Mostaan2022, Kartashov2025} and matter-wave systems \cite{Fu2022a, Fu2022b}.

{Nonlinearity brings new, unexpected features into physics of Thouless pumping because the very structure and properties of solitons differ substantially in materials with different types of nonlinearities.} This has been illustrated in recent studies of continuous single-component solitons governed by the Gross–Pitaevskii (or nonlinear Schr\"odinger) equation in one~\cite{Fu2022a, Cao2024} and two~\cite{Fu2022b} dimensions, vector \cite{Lyu2024} and multicolor optical solitons~\cite{Kartashov2025}, as well as discrete one- \cite{Jurgensen2022, Jurgensen2023} and two-component~\cite{Mostaan2022} states. In contrast to linear excitations, the pumping of solitons is strongly affected by their amplitude that allows to observe the transition between multiple regimes: the absence of pumping, quasilinear transport at sufficiently small amplitudes, transient regimes with no clear quantization, quantized transport, and nonlinear breakup of transport. Surprisingly, in all these cases, the pumping behavior is still determined by the populations and Chern numbers of the linear bands of the governing periodic Hamiltonian. This universality reflects the fact that, despite their nonlinear nature, the existence of solitons is inherently connected to the band-gap structure of the underlying linear system.

All previously reported experiments on Thouless pumping in atomic and optical systems utilized dynamically varying potentials that typically consist of two or more mutually sliding sublattices, resulting in potentials with nonzero space-time Chern numbers. A principally new situation arises when a system is subject to a periodic potential and periodic gauge field that move relative to each other. An example of such a platform is offered by spin-orbit-coupled  Bose-Einstein condensates (BECs)~\cite{Lin2011, Galitski2013}, where solitons can form~\cite{Achilleos2013, Xu2013, Kartashov2013, Sakaguchi2014, Salerno2015, Zhang2015, Salerno2016}, and where various potential and synthetic spin-orbit-coupling (SOC) landscapes can be created \cite{Ruseckas2005,Goldman2014, Jimenez2015}. These two types of lattices have fundamentally different effects, as the optical lattice affects each condensate component individually, while the SOC lattice provides a spatially inhomogeneous coupling between them. It has been previously shown that a stationary helicoidal SOC assisted by Zeeman splitting supports the existence of solitons~\cite{Kartashov2017}, while incommensurate periodic SOC and optical lattices enable the localization of cold atoms~\cite{Zezyulin2022}.

In this Letter, we reveal new mechanism of linear and nonlinear quantized Thouless pumping for two-component elongated BECs with helicoidal SOC moving with respect to a static optical lattice, identical for both components. We show that the spectral bands of the Hamiltonian with moving helicoidal SOC and lattice potential generally have nonzero space-time Chern indices which determine the quantized displacement of the center of mass of condensate over one pumping cycle in the regime of topological pumping. This displacement is accompanied by an integer number of oscillations of the pseudospin. Stable quantized transport is also obtained in interacting condensates for solitons from semi-infinite and finite spectral gaps. Depending on the chemical potential, there also exist other
pumping scenarios: arrest of pumping and broken pumping due to instabilities. In addition, we establish the significant role of Zeeman splitting in the observation of quantized transport.

\paragraph{{The model.}}
Consider the vector Gross-Pitaevskii equation (GPE) for the two-component order parameter $\bPsi=(\Psi_1,\Psi_2)^T$:   
\begin{align}
	\label{GPE}
	i\partial_t \bPsi=H\bPsi+g\left(\bPsi^\dag\bPsi\right)\bPsi,
\end{align}
where the linear Hamiltonian is given by 
\begin{align}
	\label{hamilt}	
    H=\frac{1}{2}\left[i \partial_x-A(x-vt)\right]^2+\frac{\Delta_1}{2}\sigma_1 +\frac{\Delta_3}{2}\sigma_3	+V(x),
\end{align}
$\sigma_{1,2,3}$, are the Pauli matrices, and $V(x)=V_0\cos(2px)$ is an optical lattice (OL) of the amplitude $V_0$, $p$ is an integer number. This scaling fixes the units in which coordinate and energy are measured: they are $p/k_L$ and $E_L/p^2$ where $k_L$ is the wavenumber and $E_L=\hbar^2k_L^2/(2m)$ is the recoil energy: in the case of $^{87}$Rb and $k_L\approx 3\,\mu\text{m}^{-1}$ one estimates $E_L\approx 3.2\times 10^{-31}\,$J  {(see e.g.~\cite{Hamner2015,Wang2023}). }

The vector potential is helicoidal:
$ 	A(\xi)=\alpha \boldsymbol{\sigma}\cdot \mathbf{n}(\xi)$, where $ \mathbf{n}(\xi)=\left(\cos(2q\xi),\sin(2q\xi),0\right),
$ 
$\xi=x-vt$, and $0<v\ll 1$ is the velocity of the SOC lattice. The spatial periods of both lattices are commensurate, i.e.,  $p$ and $q$ are coprime integers.   {A helicoidal SOC can be created by a linear-momentum mismatch between the counter-propagating Raman beams coupling states $m_F=0,\pm 1$ from the electronic manifold $F=1$ with excited states. For experimental realization of spatially inhomogeneous and time-dependent SOC see e.g. Refs.~\cite{Beeler2013,Olson2014}  and ~\cite{Lin2011b}. Other possibilities based on dark states of, say, tripod atomic scheme~\cite{Ruseckas2005} can be used too. Alternatively, by changing to the frame moving with velocity $v$, Eq.~(\ref{GPE}) is reduced to the model with constant SOC and moving OL similar to the one studied experimentally in~\cite{Hamner2015}.  The SOC strength $\alpha$
is tunable up to several recoil energies, through change of the intensities or angle between Raman beams.}

The coefficient $g$ in Eq.~(\ref{GPE}) describes the strength of the inter- and intra- species interactions which are assumed to be equal: $g=1$ ($g=-1$) corresponding to repulsive (attractive) interactions, while $g=0$ describing a non-interacting condensate. Since the quantization of transport is governed by the single-particle Hamiltonian $H$, we neglect the differences in interactions between the pseudospin components {(see e.g.~\cite{Lin2011,Qu2013} for rubidium atoms and \cite{Zhang2012} for theoretical study of a more general case). For our $|g|=1$, transverse trap frequency $\omega_r/2\pi\approx 300\,$Hz, and scattering length $a_s=5.29\,$nm  ($^{87}$Rb) the normalization $(\bPsi,\bPsi)=N$ [hereafter $( \mathbf{f},\mathbf{g} )=\int_{-\infty}^\infty \mathbf{f}^\dag \mathbf{g} dx$] corresponds to $\approx 400 N/p$ atoms.}

The parameters $\Delta_1$ and $\Delta_3$ in (\ref{hamilt}) characterize strengths of the longitudinal and transverse components of the Zeeman field. Their impact on  quantized transport considered here stems from the fact that the unitary stroboscopic transformation $\cH=\cS^{-1}H\cS$ with $\cS=e^{-iqvt\sigma_3}$ yields the instantaneous Hamiltonian
\begin{align}
    \label{H_gauged}
    \cH=\frac{1}{2}\left[i\partial_x-A(x)\right]^2 +V(x) + \frac{\Delta_3}{2}\sigma_3+\frac{\Delta_1}{2}A(-vt).
\end{align}
In the absence of the $x$-component of the Zeeman field (i.e., at $\Delta_1=0$), the   Hamiltonian $\cH$ is static. Thus, the helicoidal SOC cannot result in quantized transport, unless $\Delta_1 \ne 0$. On the other hand, the transverse Zeeman field, $\Delta_3$, provides additional control over the gap between two lowest bands. A sufficiently wide gap inhibits nonadiabatic effects caused by inter-band transitions and alleviates the restriction on the smallness of the velocity $v$. Additionally, this is a prerequisite for the existence of robust finite-gap solitons.
 
\paragraph{{Linear pumping.}} For any coprime integers $p,q$, the Hamiltonian (\ref{hamilt}) is periodic in space with the period $X=\pi$. It is also periodic in time with the period $T_q=\pi/(qv)$. The quantized transport is determined by the Chern numbers $C_\nu$ of the bands of Hamiltonian (\ref{hamilt}). To compute them we solve the instantaneous linear eigenvalue problem $H\bvarphi_{\nu k}=\mu_{\nu k}\bvarphi_{\nu k}$, where $t$ is treated as a parameter, $\mu_{\nu k}(t)$ is the eigen-energy corresponding to the Bloch function (BF) $\bvarphi_{\nu k}(x,t)=e^{ikx}\bu_{\nu k} (x,t)$, $\bu_{\nu k} (x,t)=\bu_{\nu k} (x+X,t)$ is its spatially periodic part, $\nu= 1,2,\ldots$ enumerates the Bloch bands, and $k$ is the Bloch wavenumber in the reduced Brillouin zone (BZ). The Chern number for band $\nu$ is defined as $C_\nu = \pi^{-1}{\rm Im}\int_0^{T_{q}}dt\int_{BZ} dk  \langle \partial_k \bu_{\nu k}, \partial_t \bu_{\nu k} \rangle$, where $\langle \mathbf{f},\mathbf{g}\rangle=\int_{0}^X \mathbf{f}^\dag \mathbf{g} dx$.
    
\begin{figure}
    \includegraphics[width=0.99\columnwidth]{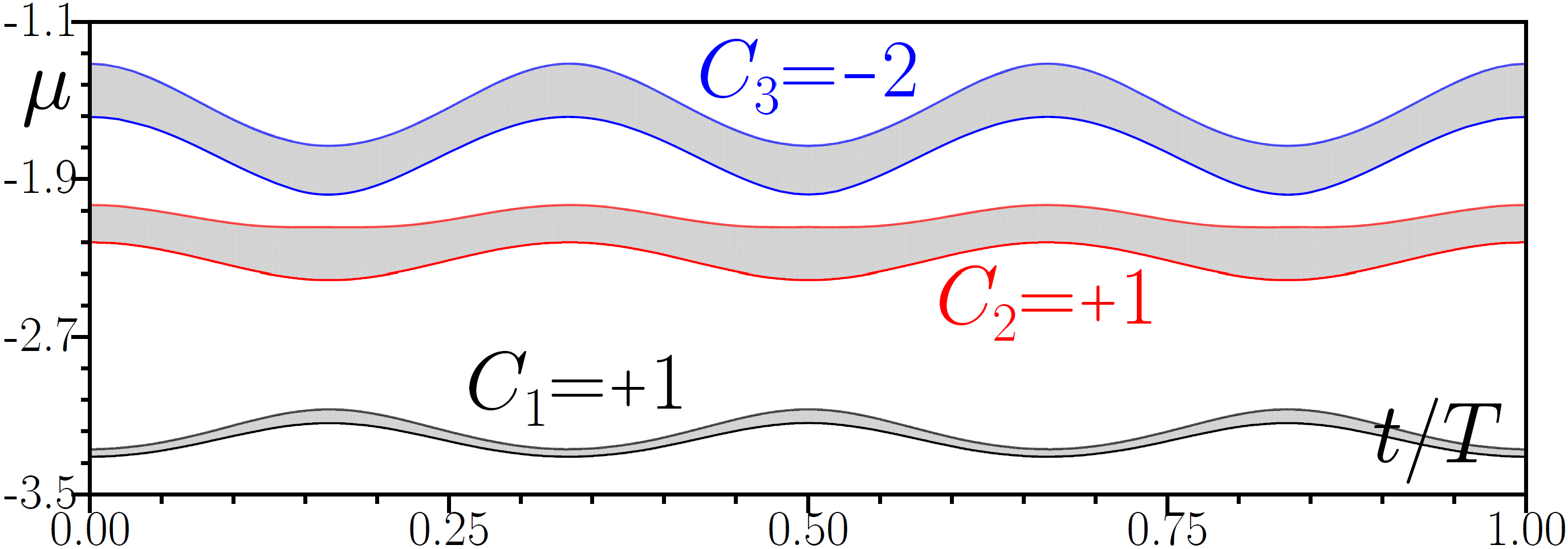}%
 \caption{Evolution of band edges of the Hamiltonian $H$ with time $t$ on one period $T$. Space‐time Chern numbers are indicated near corresponding bands. Hereafter we use the parameters $p=3$, $q=1$, $\alpha=6$, $V_0=6$, $\Delta_1=4$, and $\Delta_3=8$.}
 	\label{fig:bands}
 \end{figure}

Not every choice of coprime integers $(p,q)$ results in topologically nontrivial bands with $C_\nu \ne 0$. In particular, all Chern numbers are zero for $p=q$. For $p<q$ the simplest topologically nontrivial combination is $(p,q)=(2,3)$. In this case the total period $X=\pi$ of the Hamiltonian (\ref{hamilt}) is larger than OL and SOC periods. For $p>q$ nonzero $C_\nu$ are obtained for $(p,q)=(2,1)$, $(3,1)$, etc. Below we use the case $(p,q)=(3,1)$ because in this case the period of SOC coincides with $X$. One-cycle evolution of three lowest bands of $H$ for this case is illustrated in Fig.~\ref{fig:bands}, where corresponding Chern numbers are indicated for each band. The eigenvalues $\mu_{\nu k}(t)$ are $T/3$-periodic ($T=T_{q=1}=\pi/v$). This additional periodicity is a consequence of a symmetry of the Hamiltonian $H$. Indeed, the shift $t\to t+ m_1T_q+m_2T_p$ [here $m_1$ and $m_2$ are arbitrary integers, $T_p=\pi/(pv)$, $T_q=\pi/(qv)$] in the instantaneous Hamiltonian results in the spatial shift of the arguments of the BFs: $x\to x +\pi m_2/p$. Thus, by choosing $m_2=p$ we obtain that $\mu_{\nu k}(t)=\mu_{\nu k}(t+T_p)$ (in Fig.~\ref{fig:bands} $p=3$ and $T_p=T/3$). This effect has direct implications for the pumping dynamics. Indeed, the dynamics of a spinor macroscopic wavefunction can be characterized by the motion of its center of mass defined as $x_c(t)= (\bPsi,x\bPsi)/N$, 
as well as by the average pseudospin $\bS = (S_1, S_2, S_3)$, where $ S_j = ( \bPsi,\sigma_j \bPsi )/N$ (hereafter $( \mathbf{f},\mathbf{g} )=\int_{-\infty}^\infty \mathbf{f}^\dag \mathbf{g} dx$). 
Since spatial shift of $\bPsi$ does not affect $\bS$, from the above arguments one concludes that $\bS(t)=\bS(t+T_p)=\bS(t+T_q)$.

 \begin{figure}
 		\includegraphics[width=0.999\columnwidth]{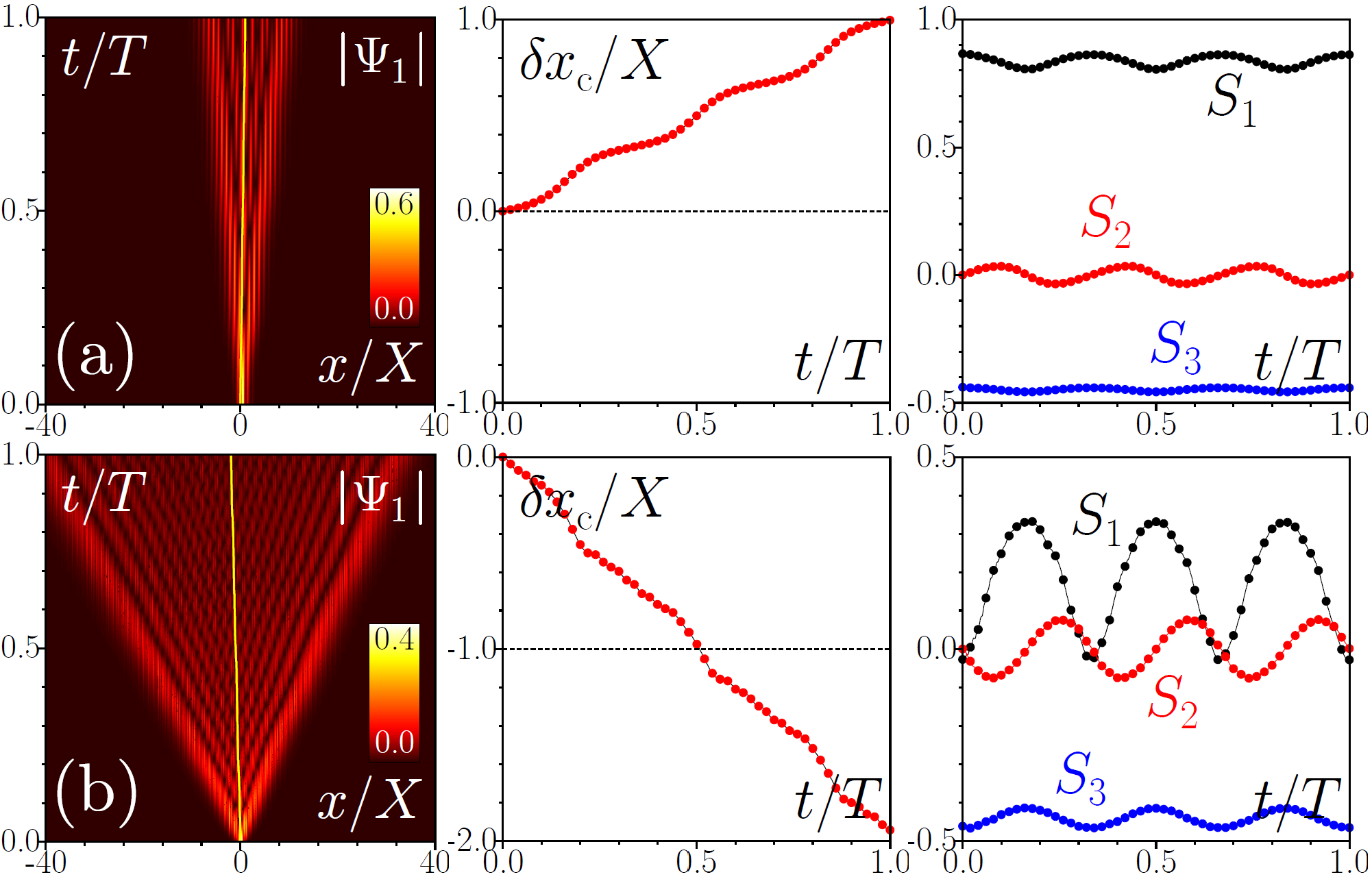}%
\caption{Evolution dynamics (first column), wavepacket displacement $\delta x_c$ vs time $t$ (second column), and evolution of pseudospin components (third column) for excitation of the first (a) and third (b) bands in the linear lattice. Here and below $v=10^{-2}$. We show evolution of only the first component $|\Psi_1|$, since the behavior of $|\Psi_2|$ is qualitatively similar.  
 }
 	\label{fig:lin}
 \end{figure}

For a non-interacting condensate (with $g=0$), we simulated the evolution dynamics  using Eq.~(\ref{GPE}) with  initial conditions uniformly exciting each of three lowest bands. For each band, the initial condition was the respective Wannier function. In each case we computed the displacement $\delta x_c(t) = x_c(t) - x_c(0)$ over one cycle of pumping.   Quantization of transport means  $\delta x_c(T) = C_\nu X$. This equality has been  confirmed in our simulations for each considered band. Results for the first and third bands are shown in Fig.~\ref{fig:lin}.  For each band the adiabatic pumping   is accompanied by appreciable {dispersion} which, as expected, increases with the band index.
Nevertheless,  the displacement   $\delta x_c(t)$  remains almost perfectly quantized after one cycle of pumping (the second column of Fig.~\ref{fig:lin}).  The pseudospin dynamics, shown in the third column of Fig.~\ref{fig:lin}, feature $T/3$-periodicity, which also agrees with the above arguments.

 \begin{figure}
    \includegraphics[width=0.999\columnwidth]{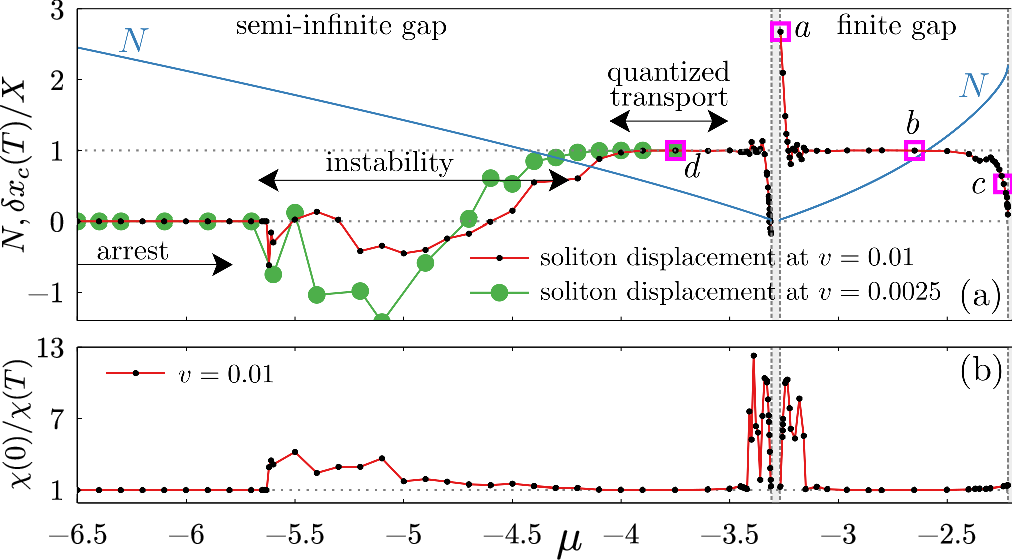}%
 \caption{{(a) Families of solitons $N(\mu)$ in the semi-infinite and finite gap for  attractive and repulsive interactions, respectively, and normalized one-cycle  displacement, $\delta x_c(T)/X$, vs $\mu$. Vertical gray stripe indicates the first spectral band. Small squares with labels $a,b,c,d$ correspond to solitons whose transport is shown in Fig.~\ref{fig:solitons}. (b) Ratio between IPRs of the initial soliton and of solution after one pumping cycle.}}
 	\label{fig:families}
 \end{figure}

 \begin{figure*}
 		\includegraphics[width=0.99\textwidth]{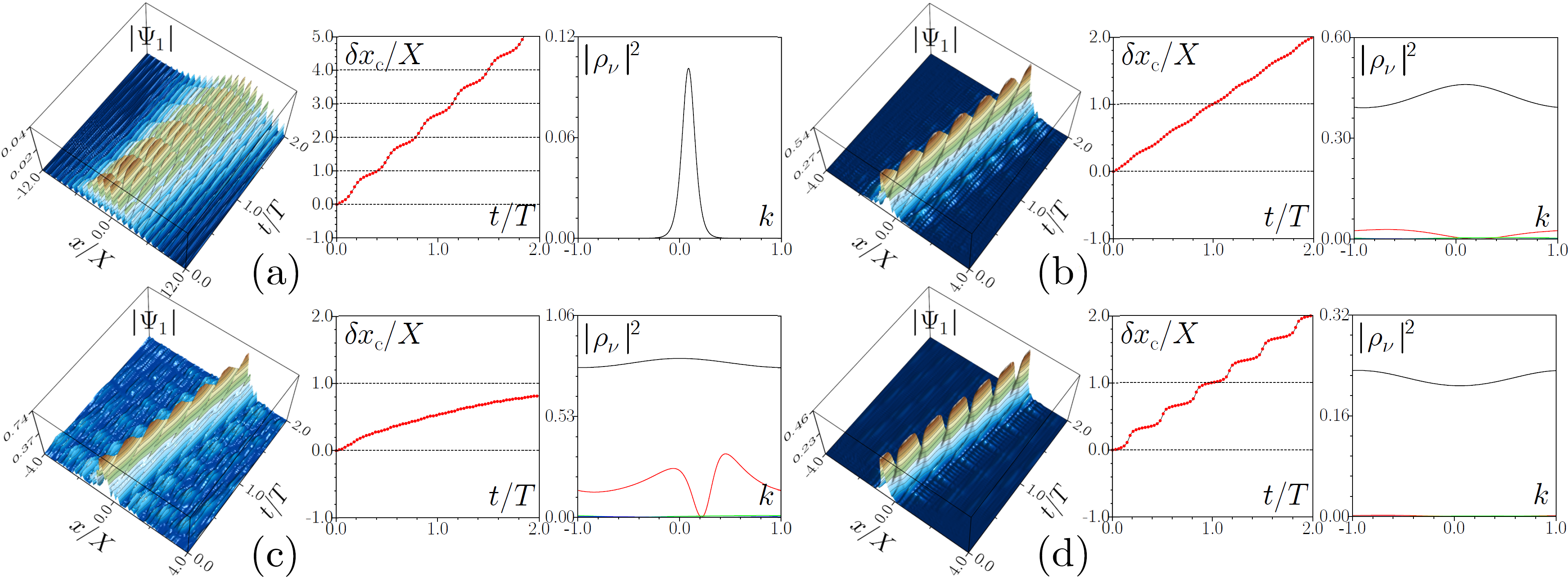}%
\caption{Topological pumping of solitons with $\mu=-3.267$ (a), $\mu=-2.650$ (b), and $\mu=-2.240$ (c) for repulsive interactions, and soliton with $\mu=-3.750$ (d) for attractive interactions {[see small squares in Fig.~\ref{fig:families}(a) highlighting the locations of these solitons]}. For each soliton, we plot the dynamics of the first component ($\Psi_1$) over two pumping cycles at $v=10^{-2}$, displacement of the center of mass, and projection of $\bPsi$ on Bloch bands at $t=0$ (black, red, and green lines correspond to projections on the first, second, and third bands, respectively). Only the dynamics of $|\Psi_1|$ is shown, as $|\Psi_2|$ behaves similarly.}
 	\label{fig:solitons}
 \end{figure*}

\paragraph{{Transport of solitons.}} We proceed to nonlinear transport by including the interatomic interactions. We have computed families of stationary solitons at $t=0$ corresponding to the substitution $\bPsi = e^{-i\mu t} \bpsi(x)$, where $\mu$ is the chemical potential. For attractive interactions ($g=-1$), we consider $\mu$ in the semi-infinite gap, while for repulsive interactions ($g=+1$) solitons exist in the finite gap between first and second bands. Figure~\ref{fig:families}(a) plots  the numbers of atoms $ N=(\bpsi,\bpsi)
$ in stationary solitons   against  $\mu$. Since the robust nonlinear transport is possible only for dynamically stable solutions, we have confirmed stability of the   solitons in the framework of the GPE with the time-independent Hamiltonian (obtained by setting $v=0$). Further, we have used the stationary wavefunctions $\bpsi(x)$ as initial conditions for the GPE (\ref{GPE}) and  simulated the dynamics of the condensate with moving   SOC. Figure~\ref{fig:families}(a) plots the one-cycle displacement  against   $\mu$. Quantized transport is possible for solitons in both finite and infinite gaps, within certain ranges of the chemical potential. At the same time, we observe several other scenarios, including the  region of instability (where the adiabatically moving SOC lattice destroys soliton structure), as well as the arrest of  transport for large-amplitude  solitons in the semi-infinite gap.

{Figure~\ref{fig:families}(b)  characterizes the dispersive broadening of solitons after one pumping cycle using the inverse participation ratio  (IPR), defined as $\chi(t) = N^{-2}\int_{-\infty}^\infty  (|\Psi_1|^4 + |\Psi_2|^4)dx$.  Small and large values of this widely used characteristic indicate weak and strong localization, respectively  \cite{Evers}.  For solitons with small $N$ the ratio $\chi(0)/\chi(T)$  is highly irregular and much larger than unity, indicating appreciable broadening at $t=T$.  However, in the regime of quantized transport  $\chi(0)/\chi(T) \approx 1$, which signifies that the soliton restores its shape after one cycle. This behavior starkly contrasts with the   dispersive broadening of linear transport.}
 
Figure~\ref{fig:solitons} illustrates pumping for initial conditions corresponding to three different solitons in the finite gap and one soliton in the semi-infinite gap. For each solution we present the evolution dynamics of the condensate wavefunction $\Psi_1$,  the evolution of the soliton displacement over two pumping cycles, and the  squared amplitudes of projections of the initial condition onto the spectral bands computed as $\rho_\nu(k) = ( \bvarphi_{\nu k}(x, 0), \bPsi(x, 0) )$. Figure~\ref{fig:solitons}(a) illustrates a small-amplitude soliton with $\mu$ close to the left edge of the finite gap. Near this gap edge, the soliton is an envelope of the linear Bloch function from which it bifurcates. Hence its projection onto the first band features a narrow peak,  while the rest of the BZ remains unpopulated. As a result, the dynamical evolution of this soliton is not quantized, {while the soliton exhibits slow broadening in the course of evolution}. As the chemical potential increases towards the center of the finite gap, the stationary solitons become narrower, and their projections onto the first spectral band nearly uniformly occupy the BZ. In this regime, the quantized transport of gap solitons becomes possible, as exemplified in Fig.~\ref{fig:solitons}(b), where  the soliton survives multiple pumping cycles (only two cycles are shown in  Fig.~\ref{fig:solitons}(b) for better visibility). As the soliton family approaches the right edge of the finite gap, the transport becomes suppressed and   features appreciable radiation, as shown in Fig.~\ref{fig:solitons}(c).  

The quantized transport is also possible for attractive interactions, for solitons in the semi-infinite gap, as shown in Fig.~\ref{fig:solitons}(d). However, for sufficiently large numbers of particles, we observe the arrest of the transport in the semi-infinite gap, as indicated in Fig.~\ref{fig:families}(a). In this regime,  solitons remain almost perfectly immobile for all times. Solitons with intermediate numbers of atoms undergo strong distortion in the course of evolution. This behavior resembles the dynamical instability which is induced by the {sliding SOC} (as the solitons are stable when the sliding velocity is zero). This  instability [the corresponding interval is indicated in Fig.~\ref{fig:families}(a)] results in an irregular dependence of  $\delta x_c(T)$ on $\mu$. 

Figures~\ref{fig:solitons}(b,d)  indicate that, due to the additional periodicity of the Bloch  spectra, the spatial displacement features additional quantization which  results in the relation $\delta x_c(T/3) = C_\nu X/3$. {They also  suggest that in the nonlinear regime robust quantized transport is possible only if two conditions are satisfied: (i) The projection of the initial soliton onto the linear Bloch functions primarily overlaps with a single  band possessing a nonzero
Chern number; (ii) This projection is nearly uniform across the BZ. At the same time, these requirements are not sufficient, as the  instability and arrested transport in the semi-infinite gap occur even for solitons whose spectral projections are similar to that plotted in Fig.~\ref{fig:solitons}(d). Therefore, these regimes are manifestations of nonlinearity and cannot be fully explained in terms of projections of soliton wavefunctions on the  Bloch bands. In addition, we have repeated a part of our simulations with the  velocity reduced by four times. The obtained  soliton displacement [see Fig.~\ref{fig:families}(a)] shows that such reduction in the velocity only marginally extends the interval of the quantized transport.}

\begin{figure} 		\includegraphics[width=0.999\columnwidth]{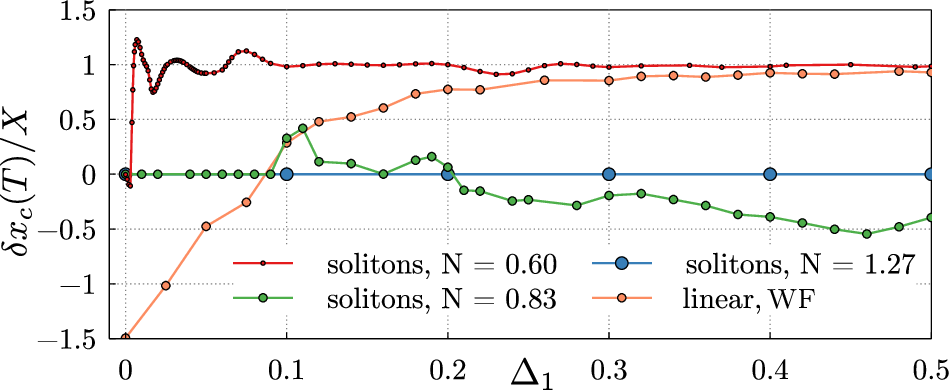}%
 	\caption{The one-cycle displacement for  solitons    (different $N$) and   a Wannier function (WF)   against the  longitudinal Zeeman field strength $\Delta_1$. 
    }
 	\label{fig:Delta1}
 \end{figure}

\paragraph{{Control of the transport by the magnetic field.}} 
At $\Delta_1=0$ the SOC lattice can be gauged out, and the quantized transport disappears both for linear waves and solitons [see Eq.~(\ref{H_gauged})]. Figure~\ref{fig:Delta1}  plots the one-cycle displacement  against  $\Delta_1$ for initial conditions in the form of solitons from the semi-infinite gap with different numbers of atoms $N$. These dependencies are juxtaposed with an analogous dependence for a non-interacting condensate, where for each value of $\Delta_1$ the initial condition was taken in the form of the Wannier function of the lowest spectral band. For solitons, we observe several representative regimes as $\Delta_1$ increases, which occur for different numbers of particles. For $N=0.60$ we encounter the onset of the quantized transport which occurs  much faster (i.e., for smaller values of the Zeeman field) than the analogous process in the non-interacting condensate. On the other hand, for sufficiently large number of particles ($N=1.27$), the transport is arrested regardless of Zeeman field strength. In the intermediate regime ($N=0.83$) the displacement  is zero for small $\Delta_1$ and then becomes irregular due to the instability.

\emph{To conclude}, quantized as well as non-quantized pumping of solitons in BECs can be induced by the sliding helicoidal SOC in the presence of an optical lattice and  a Zeeman splitting. Quantized transport  exists  for solitons in the semi-infinite gap and finite gaps.  Different regimes of pumping   include non-topological pumping of small-amplitude solitons, quantized transport, unstable propagation, as well as arrest of transport for large-amplitude solitons in the semi-infinite gap. {As a final note, the quantized transport reported here is potentially observable in other physical settings, such as optical waveguides with dispersive coupling~\cite{Chiang1997,Liu2011,Kartashov2015}, liquid crystal optical cavities~\cite{Rechcinska2019}, and discrete systems that emulate SOC~\cite{Salerno2015,Salerno2016}.}


\begin{acknowledgments}

  Y.V.K. acknowledges funding by the research project FFUU-2024-0003 of the Institute of Spectroscopy of the Russian Academy of Sciences. V.V.K. acknowledges financial support from the Fundação para a Ci\^encia e Tecnologia under the project 2023.13176.PEX (DOI https://doi.org/10.54499/2023.13176.PEX) and by national funds, under the Unit CFTC - Centro de F\'isica Te\'orica e Computacional, reference UID/00618/2025, financing period 2025-2029. The work of D.A.Z. was supported by the Priority 2030 Federal Academic Leadership Program. 

\end{acknowledgments}

\end{document}